\documentclass[prd,superscriptaddress,amsfonts,amssymb,amsmath]{revtex4-2}
\usepackage{bm}
\usepackage{amsfonts}
\usepackage{latexsym}
\usepackage[latin1]{inputenc}
\usepackage{graphicx}
\usepackage{amsmath}
\usepackage{palatino}
\usepackage{mathpazo}
\usepackage{textcomp}
\usepackage{float}
\usepackage{booktabs}
\usepackage{dcolumn}
\usepackage{booktabs}
\usepackage{multirow}
\usepackage{hyperref}
\hypersetup{colorlinks,citecolor=red}
\usepackage{amsmath}
\usepackage{xcolor}
\usepackage{orcidlink}
\usepackage[caption=false]{subfig}
\usepackage{commath}
\def\jnl@style{\it}
\def\aaref@jnl#1{{\jnl@style#1}}

\def\aaref@jnl#1{{\jnl@style#1}}

\def\aj{\aaref@jnl{AJ}}                   % Astronomical Journal
\def\apj{\aaref@jnl{ApJ}}                 % Astrophysical Journal
\def\apjl{\aaref@jnl{ApJ}}                % Astrophysical Journal, Letters
\def\apjs{\aaref@jnl{ApJS}}               % Astrophysical Journal, Supplement
\def\apss{\aaref@jnl{Ap\&SS}}             % Astrophysics and Space Science
\def\aap{\aaref@jnl{A\&A}}                % Astronomy and Astrophysics
\def\aapr{\aaref@jnl{A\&A~Rev.}}          % Astronomy and Astrophysics Reviews
\def\aaps{\aaref@jnl{A\&AS}}              % Astronomy and Astrophysics, Supplement
\def\mnras{\aaref@jnl{Mon.~Not.~Roy.~Astron.~Soc.}}             % Monthly Notices of the RAS
\def\prd{\aaref@jnl{Phys.~Rev.~D}}        % Physical Review D
\def\prc{\aaref@jnl{Phys.~Rev.~C}}  % Physical Review C
\def\prl{\aaref@jnl{Phys.~Rev.~Lett.}}    % Physical Review Letters
\def\qjras{\aaref@jnl{QJRAS}}             % Quarterly Journal of the RAS
\def\skytel{\aaref@jnl{S\&T}}             % Sky and Telescope
\def\ssr{\aaref@jnl{Space~Sci.~Rev.}}     % Space Science Reviews
\def\zap{\aaref@jnl{ZAp}}                 % Zeitschrift fuer Astrophysik
\def\nat{\aaref@jnl{Nature}}              % Nature
\def\aplett{\aaref@jnl{Astrophys.~Lett.}} % Astrophysics Letters
\def\apspr{\aaref@jnl{Astrophys.~Space~Phys.~Res.}} % Astrophysics Space Physics Research
\def\physrep{\aaref@jnl{Phys.~Rep.}}      % Physics Reports
\def\physscr{\aaref@jnl{Phys.~Scr}}       % Physica Scripta
\def\commat{\aaref@jnl{Comm.~Math.~Phys.}}              % Communications in Mathematical Physics
\def\science{\aaref@jnl{Science}}               % Science
\def\cqg{\aaref@jnl{Classical Quant.~Grav.}}            % Classical and Quantum Gravity
\def\jpcs{\aaref@jnl{JPCS}}                                     % Journal of Physics Conference Series
\def\ijmpd{\aaref@jnl{Int.~J.~Mod.~Phys.~D}}                    % International Journal of Modern Physics D
\def\grg{\aaref@jnl{Gen.~Relat.~Gravit.}}               % General Relativity and Gravitation
\def\rpp{\aaref@jnl{Rep.~Prog.~Phys.}}          % Reports on Progress in Physics
\def\npa{\aaref@jnl{Nucl.~Phys.~A}}        % Nuclear Physics A
\def\lrr{\aaref@jnl{Living Rev.~Rel.}}                   % Living reviews in relativity
\def\jcap{\aaref@jnl{J.~Cosmology Astropart.~Phys.}}    % Journal of cosmology and astroparticle physics
\def\rmp{\aaref@jnl{Rev.~Mod.~Phys.}}   %Reviews of modern physics
\def\epjc{\aaref@jnl{Eur.~Phys.~J.~C}}

\allowdisplaybreaks[1]
\begin{document}

\color{black}       %% For one column

\title{\bf Observationally constrained accelerating cosmological model with higher power of non-metricity and squared trace}

\author{A. S. Agrawal \orcidlink{0000-0003-4976-8769}}
\email{agrawalamar61@gmail.com}
\affiliation{Department of Mathematics,
Birla Institute of Technology and Science-Pilani, Hyderabad Campus,
Hyderabad-500078, India.}

\author{B. Mishra \orcidlink{0000-0001-5527-3565}}
\email{bivu@hyderabad.bits-pilani.ac.in}
\affiliation{Department of Mathematics,
Birla Institute of Technology and Science-Pilani, Hyderabad Campus,
Hyderabad-500078, India.}

\author{S. K. Tripathy \orcidlink{0000-0001-5154-2297}}
\email{tripathy\_sunil@rediffmail.com}
\affiliation{Department of Physics, Indira Gandhi Institute of Technology, Sarang, Dhenkanal, Odisha-759146, India.}

%%%%%%%%%%%%%%%%%%%%%%%%%% DATE %%%%%%%%%5%%%%%%%%%%%%%%%%%%%%%%
%%%%%%%%%%
%\date{\today}

\begin{abstract}
\textbf{Abstract}
In this paper, a cosmological model of the Universe is presented in $f(Q,T)$ gravity and the parameters are constrained by cosmological data sets. Initially, a generalised form of $f(Q,T)$ model is used as $f(Q,T)=-\lambda_{1} Q^{m}-\lambda_{2} T^2$, where $\lambda_{1}$, $\lambda_{2}$ and $m$ are model parameters. With some algebraic manipulation, the Hubble parameter is obtained in terms of redshift. Then, using MCMC analysis,  the model parameters are constrained using the most current Hubble and Pantheon$^{+}$ data.  The model parameters are also verified through the BAO data set. The model shows an early deceleration transitioning to an accelerating phase of the Universe. The $Om(z)$ diagnostics indicate a positive slope, favouring the model to be in a phantom field dominated phase.    
\end{abstract}

\keywords{}
\maketitle
%\textbf{PACS number}:\\
\textbf{Keywords}: $f(Q,T)$ gravity, Dark energy, $H(z)$ data, $Pantheon^{+}$ data, BAO data, $Om(z)$ diagnostic.

\section{Introduction} \label{SecI}
The late time acceleration \cite{Riess98, Perlmutter99, Tegmark04, Abazajian04, Spergel03, Parkinson12} is one of the most significant accomplishment in recent research on cosmology and gravitational physics. This finding compels to go outside the frame to explain the repulsive nature of gravity on large cosmic scales. It is not known regarding the source of creation of the repulsive gravity,  however the reason might be the presence of dark energy. The presence of dark energy, a non-standard component of the Universe with negative pressure, or a large-scale infra-red modification of General Relativity (GR). Dark energy has major share in the mass-energy budget of the Universe and remains sub-dominant in prior epochs resulting difficulty in model building. The current dynamics of the Universe show the acceleration due to the dark energy component and the cosmological constant ($\Lambda$) may be the simplest candidate. This has been originated from early vacuum quantum fluctuations. At late time of the cosmic dynamics, the concordance flat $\Lambda$CDM paradigm has become most successful in explaining the observed phenomena. The equation of state (EoS) parameter of dark energy mediated through the cosmological constant enables to distinguish different phases of the cosmological models ($\omega_{DE} \approx -1$). The other two ideas are the quintessence and phantom field dominated phases, which can be respectively identified as, $-1 \leq \omega_{DE}< 0$ and $\omega_{DE} \leq -1$. However, the recent surveys ruled out the possibility of $\omega_{DE} \approx -1$, although $\omega_{DE}$ could be a bit less than $-1$ \cite{Riess04, Eisentein05, Astier06}. Several corrections are suggested to $\omega_{DE}$:  CMB observations was limited to $\omega_{DE} = -1.073^{+0.0900}_{-0.089}$ by the 9-year WMAP survey \cite{Hinshaw13}, $\omega_{DE} = -1.0840 \pm 0.063$ is suggested by a combination of CMB and Supernova data \cite{Hinshaw13}. Kumar and Xu constrained it to $\omega_{DE} = -1.06^{+0.110}_{-0.13}$ based on a combined examination of the data sets of SNLS3, BAO, Planck, WMAP9, and WiggleZ \cite{Kumar14}. Combining Planck data with additional astronomical data, including Type Ia supernovae, Ade et al. \cite{Ade16} suggested  $\omega_{DE} = -1.006\pm 0.045$.\\  

We shall discuss here some of the recent findings of the cosmological models with the observational data sets. Wei and Zhang \cite{Wei07a} have used the $H(z)$ data set to capture the key aspects of ten cosmological models;  six of them failed to sustain with the observational data sets, however the remaining four were compatible with the data sets. Further it has been extended to eleven interacting dark energy models with varying couplings to the observable $H(z)$ data \cite{Wei07b}. However, none of these models outperforms the simplest $\Lambda$CDM model. Seikel et al. \cite{Seikel12} have studied novel consistency tests for the $\Lambda$CDM model and the parameters are expressed explicitly in terms of $H(z)$.  Magana et al. \cite{Magana14} have investigated five different dark energy models with variable equations of state as a function of redshift. Farooq et al. \cite{Farooq17} constructed an updated Hubble parameter $H(z)$ at redshifts $0.07<z<2.36$ and used it to restrict model parameters in both spatially flat and curved dark energy cosmological models.
Mukherjee and Banerjee \cite{Mukherjee17} take a kinematic method to describe late time dynamics of the Universe with constant value of the jerk parameter. Using a kinematic technique, Mamon et al. \cite{Mamon18} have investigated accelerated expansion phase of the Universe. The deceleration parameter $q$ has been parametrized in a model-independent manner. Using the Hubble data set and latest joint light curves, Amirhashchi \cite{Amirhashchi18} has further tightened the model parameters range for the $\Lambda$CDM model. Cao et al. \cite{Cao18} have used the observational $H(z)$ data acquired from cosmic chronometers and BAO methods to evaluate the validity of $\Lambda$CDM using the two-point $Omh^{2}(z_{2};z_{1})$ diagnostic. In the context of $f(Q,T)$ gravity, the best fit values of the model parameters are found using the $R^{2}$-test formula and the available observational data sets by Pradhan et al. \cite{Pradhan20}. Yadav et al. \cite{Yadav21} studied a bulk viscous Universe with dark energy dominance in Bianchi type I space-time and used observational $H(z)$ data (OHD) and a hybrid OHD and Pantheon compilation of SN Ia data to constrain the  parameters. Also, a kinematic expression for the deceleration parameter was developed, and its current value $q_{0} =-0.5820^{+021}_{-0.023}$ and transition redshift $z_{t} = 0.723^{+0.34}_{-0.16}$ were constrained. Yadav et al., \cite{Yadav21a} investigated an anisotropic model of transitioning Universe with hybrid scalar field in Brans-Dicke gravity and calculated the age of the Universe. With various data source Amirhashchi et al., \cite{Amirhashchi22} studied the coupling between dark energy and dark matter in an anisotropic Bianchi type I space-time. Using recent $H(z)$ and Pantheon compilation data, Goswami et al., \cite{Goswami21}, studied a bulk viscous anisotropic Universe and limited its model parameters. By bounding their derived model with recent $H(z)$ data, Pantheon data, and joint $H(z)$ and Pantheon data, they estimated the current value of Hubble constant as $H_{0} = 69.39\pm 1.54~ km s^{-1}M pc^{-1}$, $70.016 \pm 1.65~ km s^{-1}M pc^{-1}$ and $69.36\pm 1.42~ km s^{-1}M pc^{-1}$ using cosmic chronometric approach. Using parametrization technique Lohakare et al. have recreated a cosmological model in a modified teleparallel Gauss-Bonnet gravity \cite{Lohakare23}.\\

Another geometric method to measure the rate of expansion of the Universe is the baryon acoustic oscillation (BAO) method. During recombination, sound waves in baryon-photon plasma are frozen as density fluctuations, with the sound horizon determining a different scale \cite{Peebles70,Sunyaev70,Holtzman89}. These sound waves appear as a peak in the matter correlation function, or equivalently as a series of oscillations in the power spectrum, at the scale of the sound horizon. CMB measurements, which yield the physical matter and baryon densities that control the sound speed, expansion rate, and recombination time in the early Universe. This can be used to predict the length scale, which corresponds to the sound horizon at the $r_{s}(z_{d})$ baryon drag epoch and the most recent determination is  $r_{s}(z_{d})$ = 153.3 $\pm$ 2.0 Mpc \cite{Komatsu09}. From three separate galaxy surveys, significant BAO detections have been reported as  i) The Sloan Digital Sky Survey (SDSS)  \cite{Kessler09, Eisenstein05, Blake07, Padmanabhan07, Abazajian09, Crocce11, Padmanabhan12}, ii) The WiggleZ Dark Energy Survey (WiggleZ) \cite{Blake11}, and iii) The 6-degree Field Galaxy Survey (6dFGS) \cite{Beutler11}.  The most exact BAO observations were analysed by comparing the SDSS, particularly the Luminous Red Galaxy (LRG) component \cite{Eisenstein05}. Eisenstein et al. \cite{Eisenstein05} reported a convincing discovery of the acoustic peak in the SDSS Third Data Release (DR3) LRG sample with effective redshift $z = 0.35$ using a two-point correlation function. Beutler et al. \cite{Beutler11} have announced a BAO finding at $z=0.1$ in the low-redshift Universe by the 6dFGS. At higher redshifts, the WiggleZ Survey quantified BAOs at $z = 0.6$, producing a $\sim 4$ percent measurement of the baryon acoustic scale. \\

At the classical level, different approaches have been proposed recently to explain the observational results. However, a satisfactory explanation of the underlying gravity theory is still awaited. One of the simplest ways to extend Einstein's gravity is to incorporate an arbitrary function of the Ricci scalar $R$ into the gravitational action. As an extended approach,  a non-minimal coupling between geometry and matter is assumed in the Einstein-Hilbert action. This later method developed different classes of gravitational theories namely $f(R,\mathcal{L}_{m})$ gravity \cite{Bertolami07}, $f(R,T)$ gravity \cite{Harko11}. Altogether a different method is also adopted to understand the gravity at long range, known as the teleparallel gravity theory. The fundamental concept behind the teleparallel approach is to replace the metric tensor of space-time, the basic physical variable describing gravitational properties, with a collection of tetrad vectors. Torsion generated by tetrad fields can then be used to explain the gravitational effects, with curvature replaced by torsion. According to the above presentation, GR has two equal representations: (i) curvature representation and  (ii) teleparallel representation. In curvature representation, the torsion and non-metricity vanish whereas in teleparallel method, the curvature and non-metricity vanish. Another equivalent representation can be realised where the geometrical variable that describes the properties of the gravitational interaction can be represented as non-metricity; vanishing curvature and torsion. The non-metricity would describe the variation of the length of a vector in teleprallel transport. This  approach is known as symmetric teleprallel approach \cite{Nester99}; it also has the benefit of covariantinzing the usual general relativity coordinate calculations. It turns out that the associated energy-momentum density in symmetric teleparallelel gravity (STG) is basically the Einstein pseudotensor, which becomes a true tensor in this geometric representation. Of late, the STG theory has gained a lot of research interest. Considering higher power of non-metricity term, the model parameters have been constrined using Hubble and Pantheon$^{+}$ data sets through MCMC analysis \cite{Narawade22a}. 

Harko et al., considered an extension of STG theory by introducing, within the context of metric-affine formalism, a new class of theories in which the non-metricity $Q$ is non-minimally coupled to the matter Lagrangian \cite{Harko18}. Further with a non-minimal coupling between $Q$ and trace $T$ of energy momentum tensor, Xu et al. \cite{Xu19} have extended the STG theory to the $f(Q,T)$ gravity.  The coupling between $Q$ and $T$ leads to non-conservation of the energy-momentum tensor in $f(Q,T)$ theory. Different aspects of the cosmological models are recently studied using the $f(Q,T)$ gravity. Some cosmological models presuming some cosmic dynamics and constrained through observational data analysis provided a satisfactory explanation of the late time cosmic acceleration scenario \cite{Pati21, Agrawal21} and different Rip scenarios \cite{Pati22}. A dynamical system analysis has been carried out in the set up of symmetric teleparallel gravity \cite{Narawade22}. The transient behaviour of the model can also be observed \cite{Zia21}. Also, with $H(z)$ and Pantheon data, $f(Q,T)$ gravity model can be fitted with $\Lambda$CDM model \cite{Najera21,Godani21}. In this paper, we have presented an accelerating cosmological model of the Universe in $f(Q,T)$ gravity and parametrize the cosmological parameters with the available cosmological data sets. The paper is organised as follows: In Sec. \ref{SecII}, the field equations of $f(Q,T)$ gravity have been constructed using a generalized functional form $f(Q,T)=-\lambda_1Q^{m}-\lambda_2T^2$ i.e. considering the squared trace. In Sec. \ref{SecIII}, the observational data sets are discussed and the parametrization of Hubble parameter is done along with Pantheon$^{+}$ data set. In Sec. \ref{SecIV}, the baryon acoustic oscillation data set is used to obtain certain constraints on the model parameters. In Sec. \ref{SecV}, we analysed the behaviour of cosmographic parameters. In Sec. \ref{SecVI}, Om diagnostic study on the model is discussed. The results and discussions of the models are given in Sec. \ref{SecVII}. Through out the paper, we use the unit system $G=c=1$.

\section{Field equations of $f(Q,T)$ gravity} \label{SecII}

The $f(Q,T)$ gravity is described through the action\cite{Xu19},

\begin{equation} \label{eq.1}
S=\int\left(\dfrac{1}{16\pi}f(Q,T)+\mathcal{L}_{m}\right)d^{4}x\sqrt{-g}.
\end{equation}
$\mathcal{L}_{m}$ and $g=det(g_{\mu \nu})$ are respectively the matter Lagrangian  and the determinant of the metric tensor. The non-metricity is defined as,
\begin{equation} \label{eq.2}
Q\equiv -g^{\mu \nu}( L^k_{~l\mu}L^l_{~\nu k}-L^k_{~lk}L^l_{~\mu \nu}) , 
\end{equation}
where the disformation is $L^k_{~l\gamma}\equiv -\frac{1}{2}g^{k\lambda}(\bigtriangledown_{\gamma}g_{l\lambda}+\bigtriangledown_{l}g_{\lambda \gamma}-\bigtriangledown_{\lambda}g_{l\gamma})$. The field equations of $f(Q,T)$ gravity are obtained as\cite{Xu19}

\begin{equation}\label{eq.3}
-\frac{2}{\sqrt{-g}}\bigtriangledown_{k}(f_{Q}\sqrt{-g}P^{k}_ {\mu \nu})-\frac{1}{2}fg_{\mu \nu}+f_{T}(T_{\mu \nu}+\Theta_{\mu \nu})-f_{Q}(P_{\mu kl} Q^{\;\;\; kl}_{\nu}-2Q^{kl}_{\;\;\;\mu} P_{kl\nu})=8 \pi T_{\mu \nu},
\end{equation}

where $f_Q=\frac{\partial f(Q,T)}{\partial Q}$ and $T_{\mu \nu}\equiv-\frac{2}{\sqrt{-g}}\frac{\delta(\sqrt{-g}\mathcal{L}_m)}{\delta g^{\mu \nu}}$ be the energy momentum tensor. Also, $\Theta_{\mu \nu}\equiv g^{\alpha \beta}\frac{\delta T_{\alpha \beta}}{\delta g^{\mu \nu}}$ and $P^{k}_{\mu \nu}=-\frac{1}{2}L^{k}_{\mu \nu}+\frac{1}{4}(Q^{k}-\tilde{Q}^{k})g_{\mu \nu}-\frac{1}{4}\delta^{k}_{(\mu}Q_{\nu)}$ is the super potential of the model. Now,
 
\begin{eqnarray} \label{eq.4}
T&=& T_{\mu \nu}g^{\mu \nu}, \nonumber\\
Q_{k}&=& Q_{k}^{\;\;\mu}\;_{\mu}, \ \ \tilde{Q}_{k}=Q^{\mu}\;_{k\mu} 
\end{eqnarray}
respectively be the trace of the energy-momentum tensor and non-metricity tensor. One should note here that, in the $f(Q,T)$ gravity, there is a violation of the energy-momentum conservation \cite{Xu19}. It may be believed that, in modified gravity theories, a violation of energy-momentum conservation  leads to a suitable explanation of the late time cosmic speed up phenomena. To frame the cosmological model, we consider a spatially flat, homogeneous and isotropic Universe through the FLRW space-time, 

\begin{eqnarray}\label{eq.5}
ds^{2}=-N^{2}(t)dt^{2}+a^{2}(t)(dx^{2}+dy^{2}+dz^{2}),
\end{eqnarray}
where the lapse function is $N(t)$ and in standard case as, $N(t)=1$, and $a(t)$ is the scale factor.  The Hubble function describes the expansion rate and can be related to the scale factor as, $H(t)=\frac{\dot{a}(t)}{a(t)}$, an over dot denotes the time derivative. $\tilde{T}=\frac{\dot{N}(t)}{N(t)}$ is the dilation rate  and for the standard case it vanishes and the non-metricity becomes, $Q=6H^{2}$. We consider the matter in the form of the perfect fluid distribution, whose energy momentum tensor can be written as, $T^{\mu}_{\nu}=diag(-\rho, p, p, p)$. For the standard case, the field equations of $f(Q,T)$ gravity \eqref{eq.3} for the flat, isotropic and homogeneous space-time can be expressed in an abstract form  \cite{Xu19,Pati21} as, 

\begin{eqnarray}
p&=&-\frac{1}{16\pi}\left[f-12FH^2-4\dot{\zeta}\right],\label{eq.6} \\
\rho&=&\frac{1}{16\pi}\left[f-12F H^2-4\dot{\zeta}\kappa_{1}\right]\label{eq.7},
\end{eqnarray}
where $F\equiv f_Q=\frac{\partial f}{\partial Q}$ and $8\pi \kappa\equiv f_{T}=\frac{\partial f}{\partial T}$, $\kappa_{1}=\frac{\kappa}{1+\kappa}$ and $\zeta=FH$.  The evolution equation of Hubble's function can be obtained by adding eqns. \eqref{eq.6} and \eqref{eq.7},
\begin{equation} \label{eq.8}
\dot{\zeta}=4\pi(p+\rho)(1+\kappa).
\end{equation}
While comparing with the Friedmann equations, the effective pressure ($p_{eff}$) and effective energy density ($\rho_{eff}$) can be expressed as, 
\begin{eqnarray}
2\dot{H}+3H^2&=&\frac{1}{F}\left[\frac{f}{4}-2\dot{F}H+4\pi[(1+\kappa)\rho +(2+\kappa)p]\right]=-8\pi p_{eff},\label{eq.9} \\
3H^2&=&\frac{1}{F}\left[\frac{f}{4}-4\pi[(1+\kappa)\rho+\kappa p] \right]=8\pi \rho_{eff}. \label{eq.10}
\end{eqnarray}

Keeping in mind, the cosmological applications of the $f(Q,T)$ gravity, three forms of $f(Q,T)$ have been suggested, such as (i) $f(Q,T)=\lambda_{1} Q+\lambda_{2} T$, (ii) $f(Q,T)=\lambda_{1} Q^{m}+\lambda_{2} T$, (iii) $f(Q,T)=-\lambda_{1} Q-\lambda_{2} T^2$, where $\lambda_1$, $\lambda_2$ and $m$ are constants model parameters \cite{Xu19}. 

We consider a model with the functional $f(Q,T)=-\lambda_{1}Q^{m}-\lambda_{2}T^{2}$. From (\ref{eq.8}) one can write the equation for $\dot{H}$ as
\begin{equation}\label{eq.11}
\dot{H}+\frac{\dot{F}H}{F}=\frac{4\pi}{F}(1+\kappa)(\rho +p).
\end{equation}
$F=f_{Q}=\frac{\partial f}{\partial Q}=-\lambda_{1}mQ^{m-1}$ and $8\pi \kappa\equiv f_{T}$ $\Rightarrow$ $\kappa=-\frac{\lambda_{2}(3\omega-1)\rho}{4\pi}$, where the EoS parameter is denoted as $\omega=p/\rho$. From eqns. \eqref{eq.7}, \eqref{eq.8} and \eqref{eq.11} the energy density can be written as
%\begin{equation}\label{eq.12}
%\rho=\frac{f-12FH^{2}}{16\pi \left[1+(1+\omega)\kappa\right]}    
%\end{equation}
%\begin{equation}\label{eq.13}
%\rho=\frac{(2m-1)\lambda_{1}Q^{m}-\lambda_{2}(3\omega-1)^{2}\rho^{2}}{16\pi \left[1+\lambda_{2}(\omega+1)(1-3\omega)\rho/4\pi\right]}    
%\end{equation}
\begin{equation}\label{eq.14}
\rho=\frac{8\pi\left[-1+\sqrt{1+\lambda_{1}\lambda_{2}(2m-1)(1-3\omega)(\omega+5)Q^{m}/64\pi^{2}}\right]}{\lambda_{2}(1-3\omega)(\omega+5)}.
\end{equation}

Power expanding the square root term in the above equation provides $\rho\propto Q^{m}$ if the condition $\lambda_{1}\lambda_{2}(2m-1)(1-3\omega)(\omega+5)Q^{m}/64\pi^{2}\ll1$ is satisfied. As a result, we have the typical general relativistic result for $m=1$ for matter dominated case in this limit. Now,

\begin{eqnarray}\label{eq.15}
\dot{H}&=&-\frac{32\pi^{2}(1+\omega)}{\lambda_{1}\lambda_{2}(2m-1)(1-3\omega)(\omega+5)mQ^{m-1}}\nonumber \\
&&\times\left[1+\frac{2\left[-1+\sqrt{1+\lambda_{1}\lambda_{2}(2m-1)(1-3\omega)(\omega+5)Q^{m}/64\pi^{2}}\right]}{(\omega+5)}\right]\nonumber \\
&&\times \left[-1+\sqrt{1+\lambda_{1}\lambda_{2}(2m-1)(1-3\omega)(\omega+5)Q^{m}/64\pi^{2}}\right].
\end{eqnarray}
Let us now define a constant parameter, $\eta=\frac{64\pi^{2}}{\lambda_{1}\lambda_{2}(2m-1)(1-3\omega)(\omega+5)}$ and a time evolving parameter $\xi (Q)=-1+\sqrt{1+Q^{m}/\eta} $, so that we have,
%\begin{equation}\label{eq.16}
%\dot{H}=-\frac{\eta(1+\omega)}{2mQ^{m-1}}\left[1+\frac{2\left[-1+\sqrt{1+Q^{m}/\eta}\right]}{(\omega+5)}\right]\left[-1+\sqrt{1+Q^{m}/\eta}\right].
%\end{equation}
\begin{equation}\label{eq.16}
\dot{H}=-\frac{\eta(1+\omega)}{2mQ^{m-1}}~\xi (Q)~\left[1+\frac{2\xi (Q)}{(\omega+5)}\right].
\end{equation}

One may notice that for decreasing values of $\lambda_{1}$ and $\lambda_{2}$ in the positive domain, the value of $\eta$ increases. As a result,  we have the limiting value $Q^{m}/\eta\ll 1$ and $Q^{m}/\eta\neq 0$ for these given conditions. Taking power expanding square root term, we may get $\xi (Q)\simeq \frac{1}{2}\frac{Q^m}{\eta}$ and $1+\frac{2\xi (Q)}{(\omega+5)}\simeq 1$ so that  \eqref{eq.16} may be approximated to
\begin{equation}\label{eq.17}
\dot{H}+\gamma H^{2}=0,    
\end{equation}
where $\gamma=\frac{3(1+\omega)}{2m}$. On solving, we obtain
\begin{equation}
H(t)=\frac{1}{\gamma t+c_1},\label{eq.18}
\end{equation} 
where $c_1$ be the integrating constant. Subsequently we can find the scale factor as, $ a(t)=c_2\left[\gamma t+c_1\right]^{\frac{1}{\gamma}}$. Using the relationship between the scale factor and the redshift parameter, $a=\frac{1}{1+z}$, we can find the parametric form of $H(z)$ as,  
\begin{equation} \label{eq.19}
H(z)=H_0\left[\sqrt{(1+z)^{3(1+\omega)}}\right]^{\frac{1}{m}}, 
\end{equation}
where $H_0$ be the present value of the Hubble parameter. For matter dominated phase $(\omega =0)$ and dark energy dominated phase $(\omega \ \text{is constant})$ respectively, the Friedman equation describes the rate of expansion $H(z)$ at redshift $z$ with constant equation of state parameter  as,
\begin{equation}\label{eq.21}
E^2(z)=\frac{H^{2}(z)}{H_{0}^{2}}=\left[(1-A)(1+z)^{3}+A(1+z)^{3(1+\omega)}\right]^{\frac{1}{m}}.
\end{equation}  
Here $A$ defines the contribution from the dark energy sector. Subsequently, we shall constrain the parameters with the cosmological data sets.
\section{Model with Observational Data}  \label{SecIII}
We shall outline in this section, the cosmological data to be used in the problem to constrain the model parameters. The data sets are related to the expansion history of the Universe i.e describing the distance-redshift relations. Mostly we shall use the expansion rate data from early type galaxies and Pantheon$^{+}$ supernovae data.    
\begin{itemize}
    \item{\bf{Hubble Data}}: Through estimations of their differential evolution, early type galaxies provide Hubble parameter measurements. The process of compilation of such observations is known as the cosmic chronometers and from the recent result, the redshift range is $0.07<z<1.965$. For these measurements, the $\chi_{OHD}^{2}$ estimator can be constructed with $32$ data points from different data source defined in TABLE \ref{TABLE III},
\begin{equation}
\chi^{2}_{OHD}=\sum_{i=1}^{32}\frac{\left[H_{th}(z_i,\theta)-H_{obs}(z_i)\right]^{2}}{\sigma_H^{2}(z_i)}, \label{eq.22}
\end{equation} 

where $H_{th}(z_i,\theta)$ and $H_{obs}(z_i)$ respectively represents the generated Hubble parameter and the observed Hubble parameter and $\theta$ be the vector of the cosmological background parameters. $\sigma_H{(z_i)}$ denote the observational errors on the measured values $H_{obs}(z_i)$.

\item{\bf{Pantheon$^{+}$ data}}: Supernovae type Ia observation was the first to indicate the accelerated expansion of the Universe. Several new SNIa data sets have been developed and in this analysis, Pantheon$^{+}$ Supernovae data is considered that contains 1701 data points and provide the estimated value of the distance moduli $\mu_i$ in the redshift range $0.00122<z<2.26$ \cite{Brout22}. The model parameters are to be fitted by comparing the observed and theoretical value of the distance moduli. The distance moduli can be defined as, 

\begin{equation}
\mu(z,\theta)=5log_{10}[d_L(z,\theta)]+\mu_0, \label{eq.23}
\end{equation}
where $\mu_0$ is the nuisance parameter and $d_L$ is the dimensionless luminosity distance defined as, 

\begin{equation}
d_L(z,\theta)=(1+z)\int_0^z \frac{dz'}{E(z')} \label{eq.24}
\end{equation}
where $E(z)=H(z)/H_0$. Now the   $\chi^2$ estimator reads,

\begin{equation}
\chi^{2}_{Pantheon}=\sum_{i=1}^{1701}\frac{\left[\mu(z_i,\theta)_{th}-\mu(z_i)_{obs}\right]^{2}}{\sigma_{\mu}^{2}(z_i)}, \label{eq.25}
\end{equation}
where $\sigma_{\mu}(z_i)$ is the standard error in the observed value. 
\end{itemize}
\begin{figure}[H]
\centering
\minipage{0.50\textwidth}
\includegraphics[width=\textwidth]{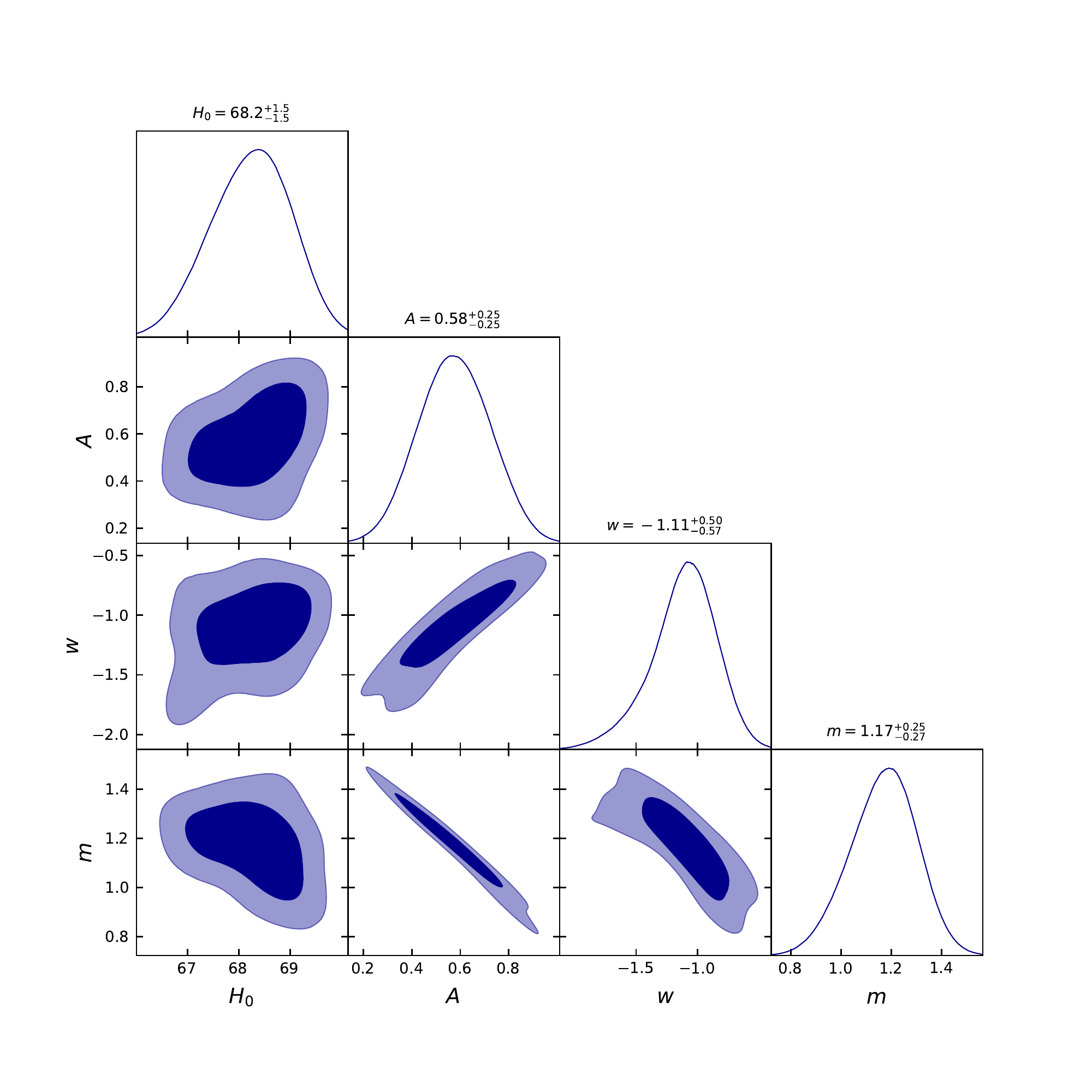}
\endminipage\hfill
\minipage{0.50\textwidth}
\includegraphics[width=\textwidth]{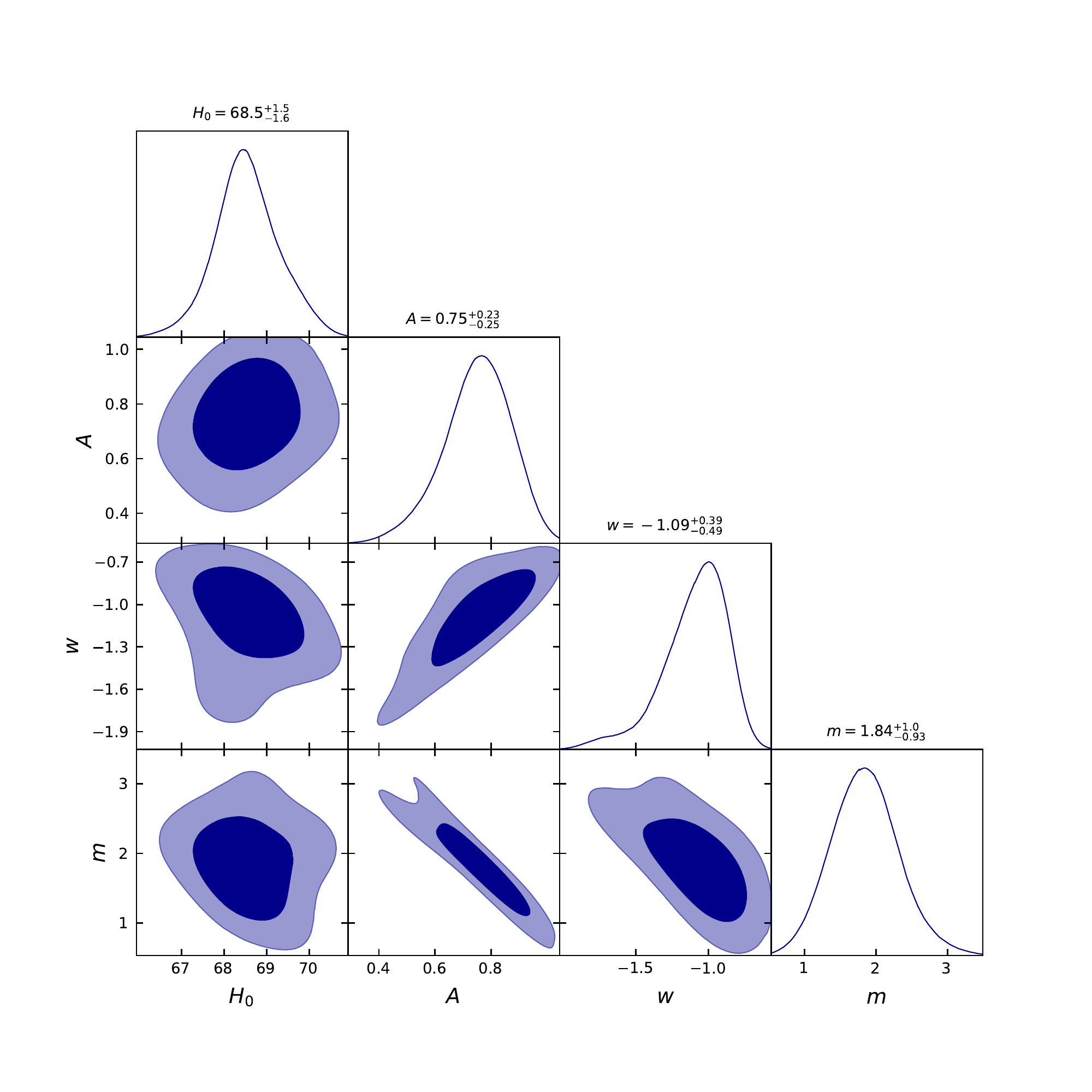}
\endminipage\hfill
\caption{Constraints on the model parameters from 32 data points of Hubble data [left panel] and Pantheon$^{+}$ data [right panel] with 1$\sigma$ and 2$\sigma$ confidence intervals.}\label{FIG.1} 
\end{figure}
%%%%%%%%%%%%%%%%%%%%%%%%%%
\begin{table} 
\caption{Summary of the observational results} % title of Table
\label{TABLE I}
\centering % used for centering table
\begin{tabular}{|l| c | c |} % centered columns (4 columns)
\hline%\hline %inserts double horizontal lines
Source/Data & Hubble & Pantheon$^{+}$   \\  % inserts table
%heading
\hline % inserts single horizontal line
$H_{0}(km s^{-1}Mpc^{-1})$ & $68.2\pm 1.5$ & $68.5^{+1.5}_{-1.6}$  \\
\hline
$\omega$ & $-1.11^{+0.50}_{-0.57}$ & $-1.09^{+0.39}_{-0.49}$  \\
\hline
$m$ & $1.17^{+0.25}_{-0.27}$ & $1.84^{+1}_{-0.93}\pm 0.054$  \\
\hline
$A$ & $0.58\pm0.25$ & $0.75^{+0.23}_{-0.25}$  \\ % [1ex] adds vertical space
\hline %inserts single line
\end{tabular}
\label{table:nonlin} % is used to refer this table in the text
\end{table}
%%%%%%%%%%%%%%%%%%%%%%%%%%%
 The model created above is compared to observational data points from $H(z)$ and pantheon$^{+}$ data. We can see the best-fitted values of the EoS parameter, $\omega$ and other parameters on the contour map for $1\sigma$ and $2\sigma$ confidence intervals. Best fit values for model parameters $m$ and $A$ are mentioned in the table \ref{TABLE I} for both Hubble and Pantheon$^{+}$ data sets. The EoS parameter becomes $\omega= -1.11$, $\omega= -1.09$ for Hubble and pantheon$^{+}$ data sets respectively.\\

%where $H_{0}$ is the current rate of expansion, $\Omega_{m}$ and $\Omega_{DE}$ respectively represent that density parameter of matter and dark energy component and $\omega(z)$ is the EoS parameter of dark energy component. 

The integral of $1/H(z)$ over redshift determines the luminosity distance. According to the analysis on the combination of low redshift data and CMB anisotropy, $30\%$ of the matter energy density ($\Omega_m$) constitutes the cold dark matter (CDM) and around $70\%$ dark energy in the form of cosmological constant $\Omega_{\Lambda}$. Also, $\Omega_m=\Omega_{CDM}+\Omega_b$ with $\Omega_{CDM}\approx0.25$ and the baryonic matter $\Omega_b\approx 0.05$. It has vanishing spatial curvature, $\Omega_{k}\approx 0$ and residual radiation, $\Omega_{r} \approx 5 \times 10^{-5}$. The radiation density parameter has been set as, $\Omega_{r} = 0$. Here, we shall discuss the simple dark energy models such as (i) $\Lambda$CDM and (ii) $\omega$CDM  with the $H(z)$ as given in Eq. \eqref{eq.21}. The $\omega$CDM approach is an extension of $\Lambda$CDM model, where  $\omega \neq -1$ exactly. Like the concordance model, a negative $\omega$ is required to ensure that the universe accelerates. Now,
%\begin{equation} \label{eq.26}
%H(z)=H_{0}\left[(1-A)(1+z)^{3}+A(1+z)^{3(1+\omega)}\right]^{\frac{1}{2m}}
%\end{equation}
\begin{figure}[H] 
\centering
\minipage{0.7\textwidth}
\includegraphics[width=\textwidth]{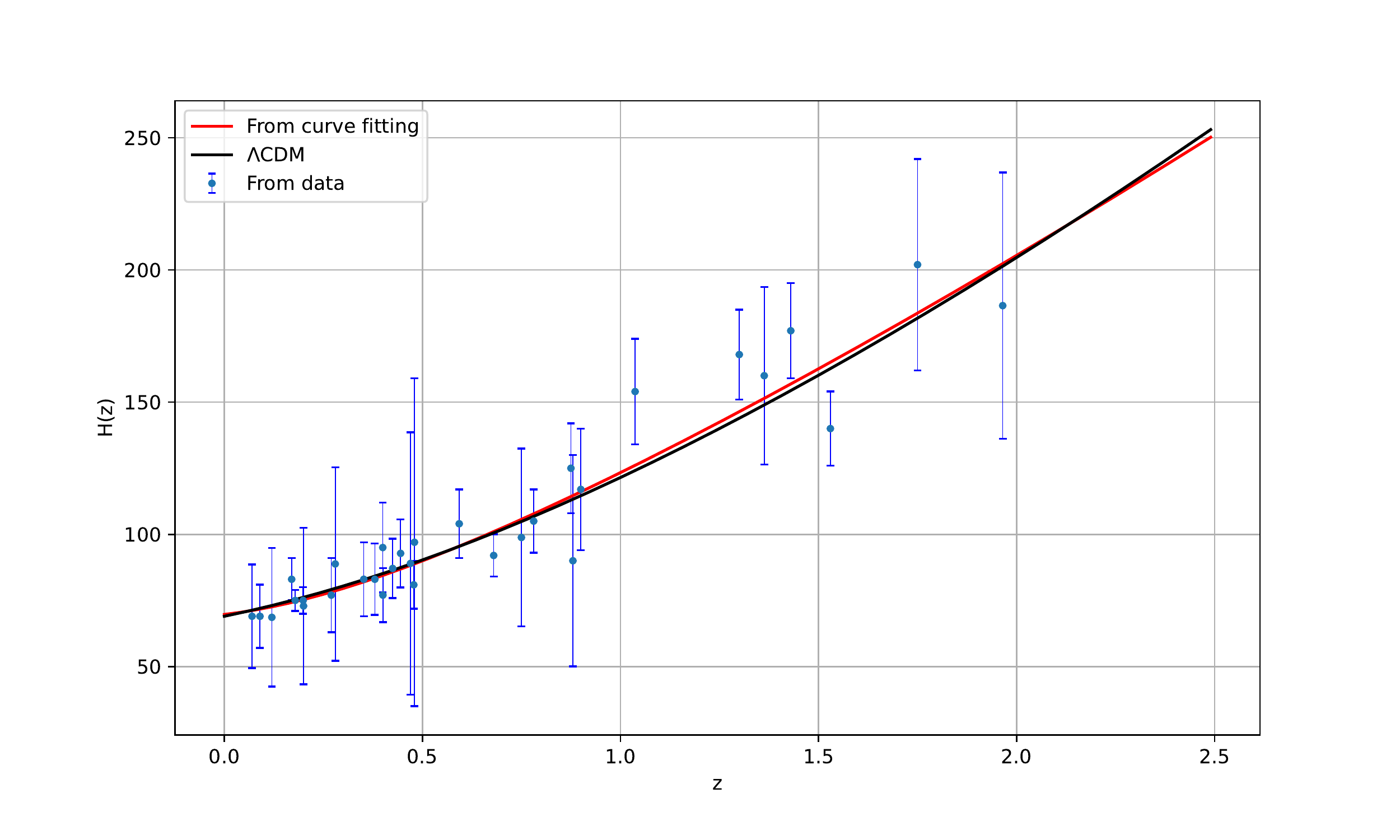}
\endminipage\hfill
\caption{Plots for $H(z)$ for derived model (red line) and $\Lambda$CDM (black line) vs redshift from 32 data points of $H(z)$ as in TABLE \ref{TABLE III}. The points with error bars indicate the observed Hubble data.} \label{FIG.2}
\end{figure} 
According to the present observations the value of Hubble parameter is $H_{0}=71\pm 3~ kms^{-1}MPc^{-1}$, and in the present scenario we constrain the value of Hubble parameter as $H_{0}\approx 68.2\pm 1.5 ~kms^{-1}MPc^{-1}$. The best fit value has been calculated using Hubble data presented in TABLE \ref{TABLE III} and FIG. \ref{FIG.2} shows that the model presented here passes almost in the middle of the observational $H(z)$ data points and it has been analysed that the presented model coincides with the well known and most appropriate $\Lambda$CDM model.

\begin{figure}[H]
\centering
\minipage{0.7\textwidth}
\includegraphics[width=\textwidth]{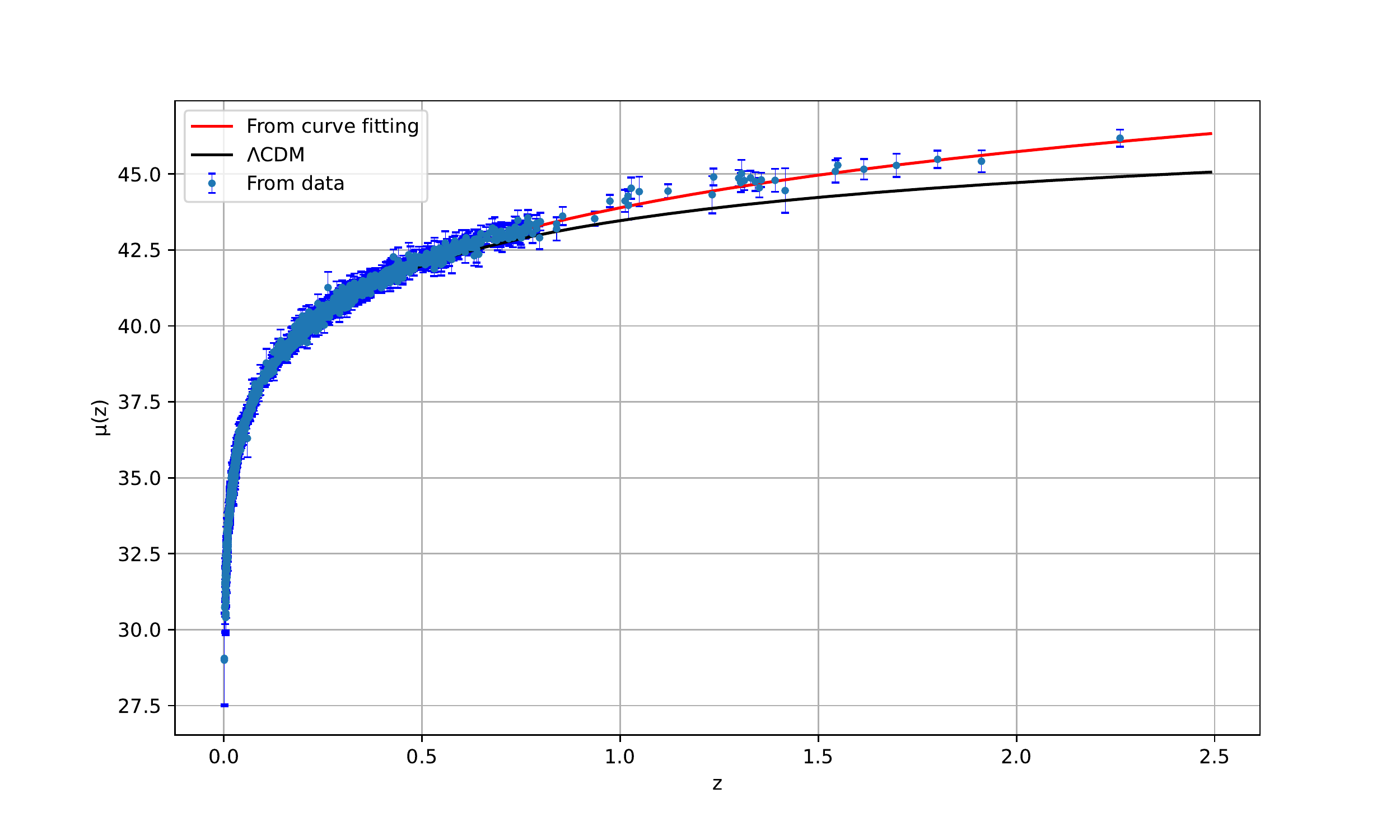} 
\endminipage
\caption{Plots of distance modulus from Pantheon$^{+}$ data.} \label{FIG.3}
\end{figure}

Now, we shall see the behaviour of distance modulus of the model with the Pantheon$^{+}$ data set. The red line in the panel of FIG. \ref{FIG.3} shows the distance modulus of the derived model and the black line represents the nature of $\Lambda$CDM model. In the plot, as compared to the $\Lambda$CDM model, the presented model fits better to the pantheon$^{+}$ data set. The recent dataset has a higher number of data points and extends longer, to $z= 2.26$ instead of $z\approx 1.5$. However, the Pantheon$^{+}$ data set is deceptive, since only six data points out of 1701 are over $z = 1.5$, and only one is above $z = 2$. The error bars have also been decreased. So, with higher number of data points, it is anticipated to get more consistent results on the recreation of the expansion history of the Universe. 

\section{Baryon acoustic oscillations and cosmic microwave background} \label{SecIV}

Numerous galaxy surveys have demonstrated that BAO standard ruler measurements are self-consistent with the standard cosmology model obtained from CMB observations, and have resulted in new, more stringent cosmological parameter constraints. The measurement of baryon acoustic oscillations (BAOs) in the large-scale clustering pattern of galaxies, as well as their use as a cosmological standard ruler, is a very promising and complementary method for mapping the distance-redshift relation \cite{Eisenstein99, Cooray01, Blake03, Linder03, Seo03}. Giostri et al. \cite{Giostri12} integrate type Ia supernovae (SN Ia) data with recent baryonic acoustic oscillations (BAO) and cosmic microwave background (CMB) measurements to limit a kink-like parametrization of the deceleration parameter. Eisenstein et al. \cite{Eisenstein07}  investigated the nonlinear degradation of the baryon acoustic signature using a number of methods. The Hubble parameter, $H(z)$, and the angular diameter distance, $D_{A}(z)$, can theoretically be derived concurrently from data in the radial and transverse directions using the baryon acoustic oscillation (BAO) scale, which gives galaxy clustering its power as a dark energy probe \cite{Blake03, Seo03, Chuang12}. As described by Jarosik et al. \cite{Jarosik11} in their WMAP data, the value of the decoupling redshift, $z_{*}$, is taken to be 1092.

In a cosmological model, $D_{V}(z)$ is a composite of the physical angular-diameter distance $D_{A}(z)$ and the Hubble parameter $H(z)$, which determine tangential and radial separations, respectively:
\begin{eqnarray}
D_{V}(z)&=&\left(\frac{D_{A}^{2}(z)cz}{H(z)}\right)^{\frac{1}{3}}, \label{eq.27}\\
D_{A}(z_{*})&=&c\int_{0}^{z_{*}}\frac{dz'}{H(z')}, \label{eq.28}\\
d_{z}&=&\frac{r_{s}(z_{d})}{D_{V}(z)}.\label{eq.29}
\end{eqnarray}

\begin{figure}[H]
\centering
\minipage{0.5\textwidth}
\includegraphics[width=\textwidth]{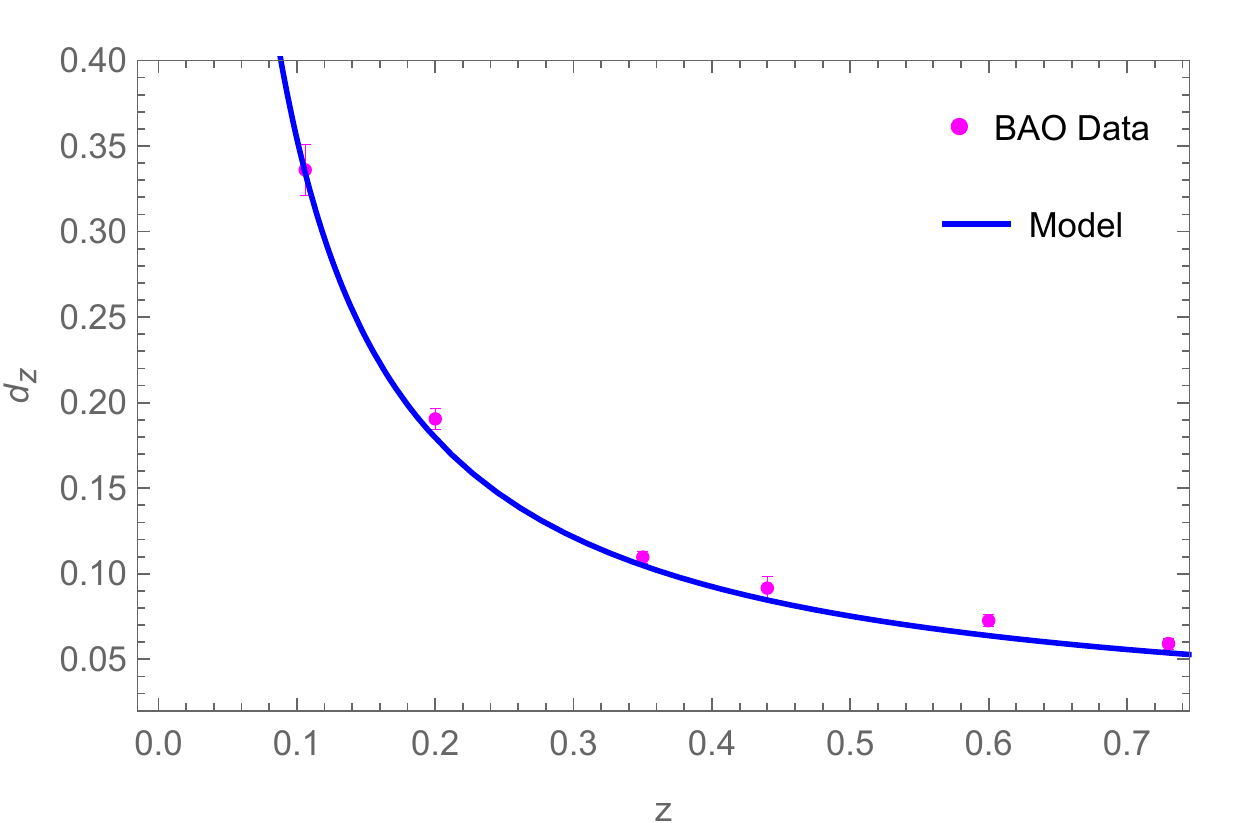}
\endminipage
\caption{Plots of $d_{z}$ parameter vs. redshift from BAO/CMB constraints,}\label{FIG.4} 
\end{figure}
%%%%%%%%%%%%%%%%%%%%%%%%%%%%%%
\begin{table}[H]
\caption{BAO measurements at six different redshifts are now included in the most recent BAO distance data set, which includes the 6dFGS, SDSS, and WiggleZ studies. Following table summarises the information.}
\label{TABLE II}
\begin{center}
\begin{tabular}{ |c|c|c|c| } 
 \hline
 Sample  & Redshift ($z$) & $d_{z}$  \\ [1ex]\hline
 6dFGS   & 0.106 & $0.3360\pm 0.0150$ \cite{Blake11}  \\ \hline
 SDSS    & 0.200 & $0.1905\pm 0.0061$ \cite{Blake11}  \\ \hline
 SDSS    & 0.350 & $0.1097 \pm 0.0036$ \cite{Blake11} \\ \hline
 WiggleZ & 0.440 & $0.0916 \pm 0.0071$ \cite{Blake11} \\ \hline
 WiggleZ & 0.600 & $0.0726 \pm 0.0034$ \cite{Blake11} \\ \hline
 WiggleZ & 0.730 & $0.0592 \pm 0.0032$ \cite{Blake11} \\ 
 \hline
\end{tabular}
\end{center}
\end{table}
%%%%%%%%%%%%%%%%%%%%%%%%%%%%
The plot for the distilled parameter has been made using the  constrained values in the model such as $H_{0}=68.5 ~kms^{-1}MPc^{-1}$, $A = 0.75$, $m = 1.84$ and $\omega = -1.09$.  As can be seen from the graphs, our results are in good agreement with the observational results of the BAO data.

\section{Cosmographic Parameters} \label{SecV}
In the cosmographic series, Hubble parameter $H$ is the first derivative form of the scale factor. The second, third and fourth derivative form of the scale factor respectively produce the deceleration, jerk and snap parameter. Determining the physical quantities $H$ and $q$ has become very crucial in characterising the evolution of the Universe. The Hubble parameter $H$ gives us the rate of expansion of the Universe so that the age of the Universe can be determined. The sign of the deceleration parameter indicates the accelerating or decelerating dynamics of the Universe, however it does not account for the entire dynamics. The change of sign of the jerk parameter $j$ is important in the sense that the positive $j$ indicates the change of expansion at some point during the evolution. The deceleration and jerk parameter define the local dynamics but can not distinguish the cosmological models effectively. When $j_{0} = 1$, the model favours the $\Lambda$CDM model and the Universe continues to expand at an accelerated rate under the influence of a cosmological constant. On the other hand, the value of $s$ is required to establish whether  there is any evolution of dark energy. The functional dependence of dark energy on the redshift $z$ is affected by departures from the predicted value of $s$, which is evaluated in the concordance model, demonstrating that it evolves as the Universe expands. The cosmographic parameters can be obtained as, 

\begin{eqnarray}
q&=&-1-\frac{\dot{H}}{H^{2}},\label{eq.30} \\
j&=&\frac{\ddot{H}}{H^{3}}-3q-2,\label{eq.31}\\
s&=&\frac{j-1}{3\left(q-\frac{1}{2}\right)}.\label{eq.32}
\end{eqnarray}

\begin{figure}[H]
\centering
\minipage{0.50\textwidth}
\centering
\includegraphics[width=\textwidth]{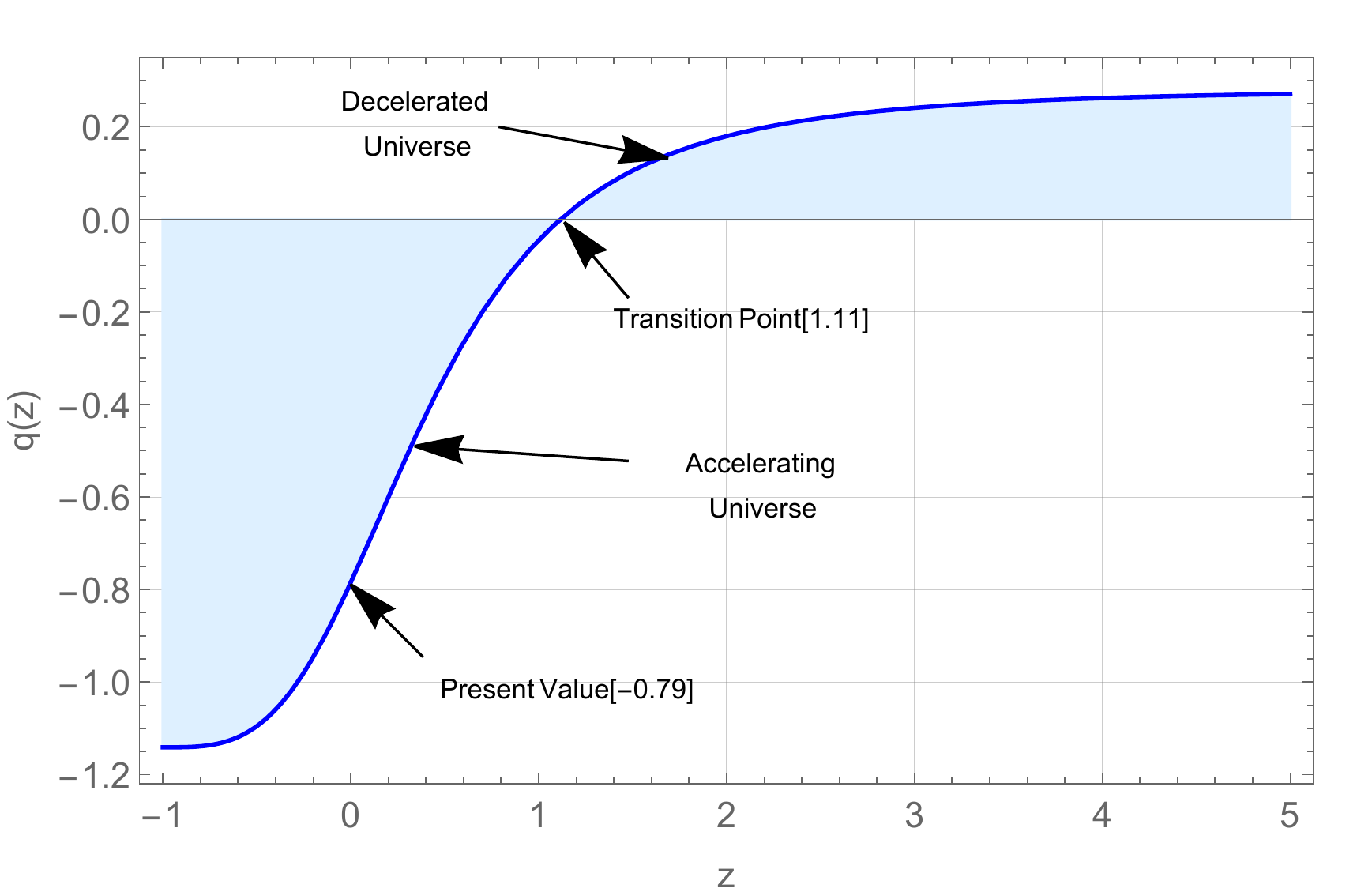}
\endminipage
\caption{Plots of deceleration parameter vs. redshift from Pantheon$^{+}$ data.}\label{FIG.5}
\end{figure}
\begin{figure}[H]
\minipage{0.50\textwidth}
\centering
\includegraphics[width=\textwidth]{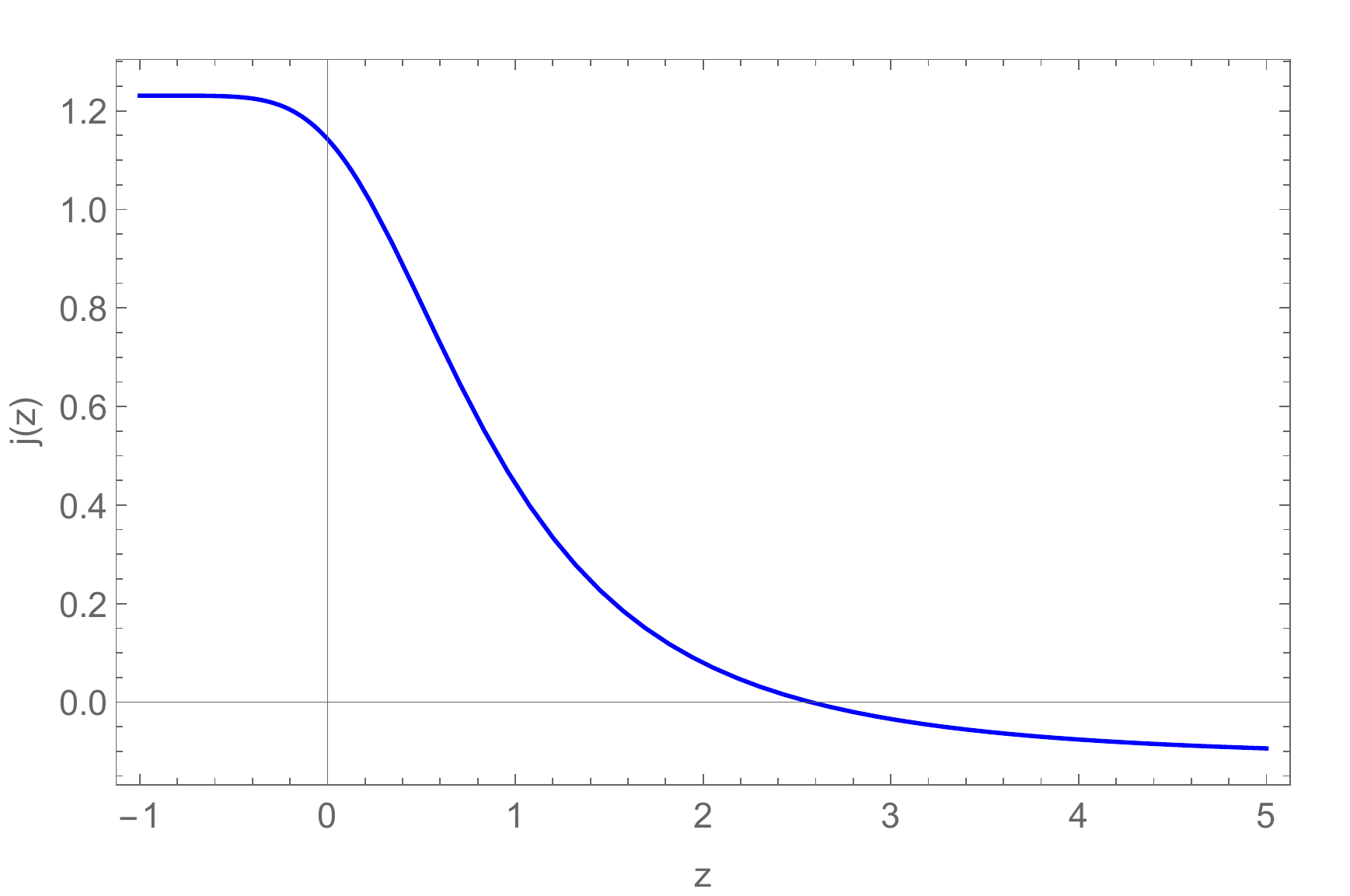}
\endminipage\hfill
\minipage{0.50\textwidth}
\centering
\includegraphics[width=\textwidth]{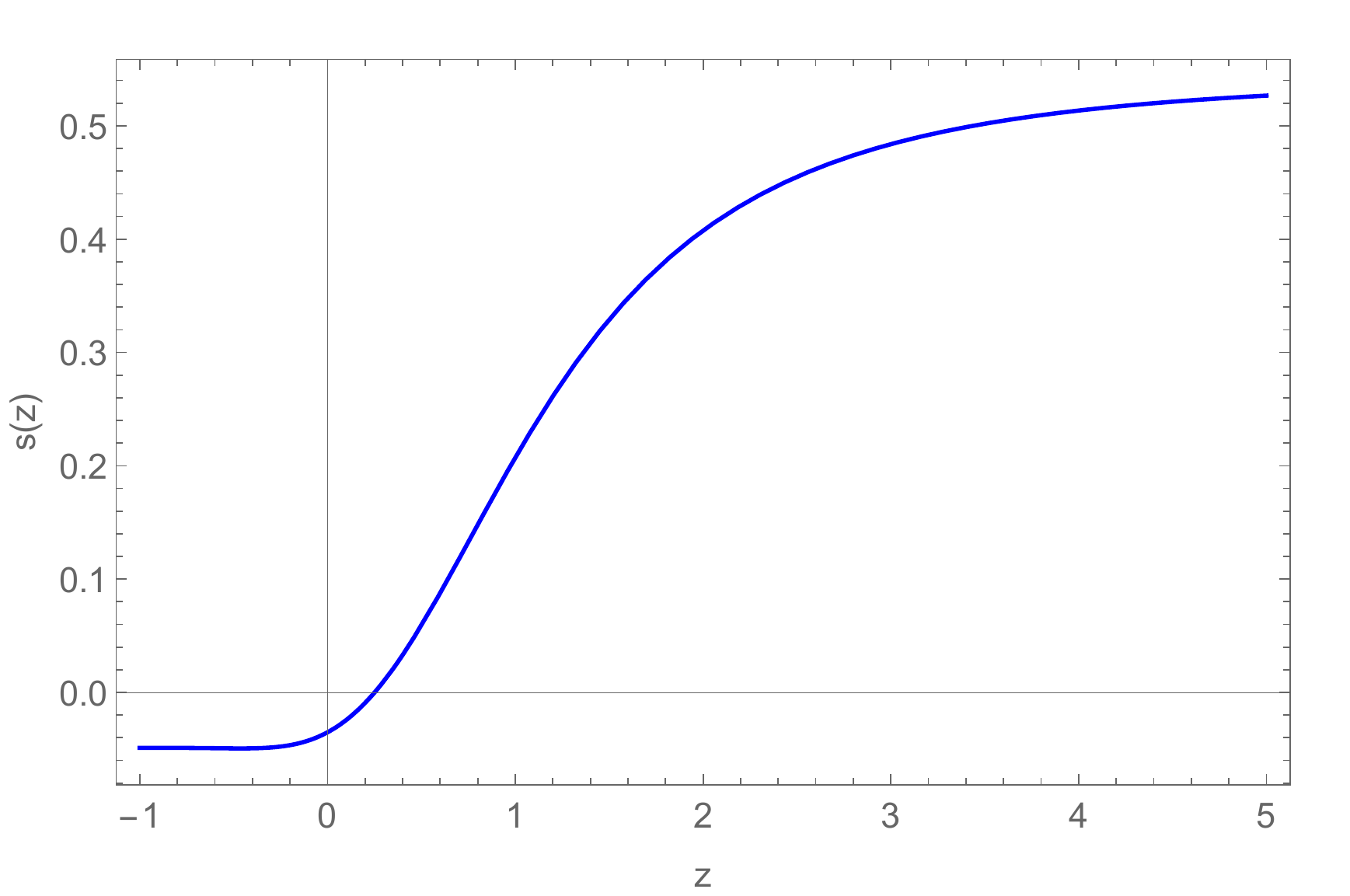}
\endminipage\hfill
\caption{Plots of jerk parameter (left panel) and snap parameter  (right panel) vs. redshift from Pantheon$^{+}$ data.} \label{FIG.6}
\end{figure}
In FIG. \ref{FIG.5}, the deceleration parameter shows early deceleration to late time acceleration, the transition occurs from deceleration to acceleration at $z_{t}=1.11$. The present value of deceleration parameter is obtained as $q \approx -0.89$. It is noteworthy to mention that the value of deceleration parameter matches with some earlier constraints,  $q=-0.56$ \cite{Capozziello20}. At late times, it approaches to $-1.18$ showing the accelerating behaviour of the model. Another important aspect of the model is its deviation from $\Lambda$CDM model, which can be identified through the state finder pair $(j,s)$ \cite{Sahni03,Alam03}. When $(j,s)=(1,0)$, the model approaches to $\Lambda${CDM} and the same feature we can obtain in this model by assuming $\omega=-1$ at present value of $z$ (see FIG. \ref{FIG.6}). However when we consider $\omega=-1.09$ (the value suggested by Pantheon$^{+}$ data), the pair approaches to $(1.21,-0.05)$ at the late time of evolution and at $z=0$ i.e for the present value of $\omega$, it gives $(j,s)=(1.14,-0.03)$. We can say that the model discussed here has an extremely similar to the $\Lambda$CDM behaviour. 

\section{Om diagnostic and Age of the Universe} \label{SecVI}
Another important cosmological diagnostic is the $Om(z)$ diagnostic that involves a combination of the Hubble parameter and cosmological redshift. This will provide a null test for cosmological constant mediated dark energy. If the value of $Om(z)$ is the same at different redshifts, then dark energy is $\approx \Lambda$, the cosmological constant. We can find the expression for $Om(z)$ as,
\begin{equation*}
Om(z)=\frac{E^{2}(z)-1}{(1+z)^{3}-1},
\end{equation*}
\begin{equation}
Om(z)=\frac{\left[(1-A)(1+z)^{3}+A(1+z)^{3(1+\omega)}\right]^{\frac{1}{m}}-1}{(1+z)^{3}-1}.\label{eq.33}
\end{equation}
\begin{figure}[H]
\centering
\minipage{0.50\textwidth}
\includegraphics[width=\textwidth]{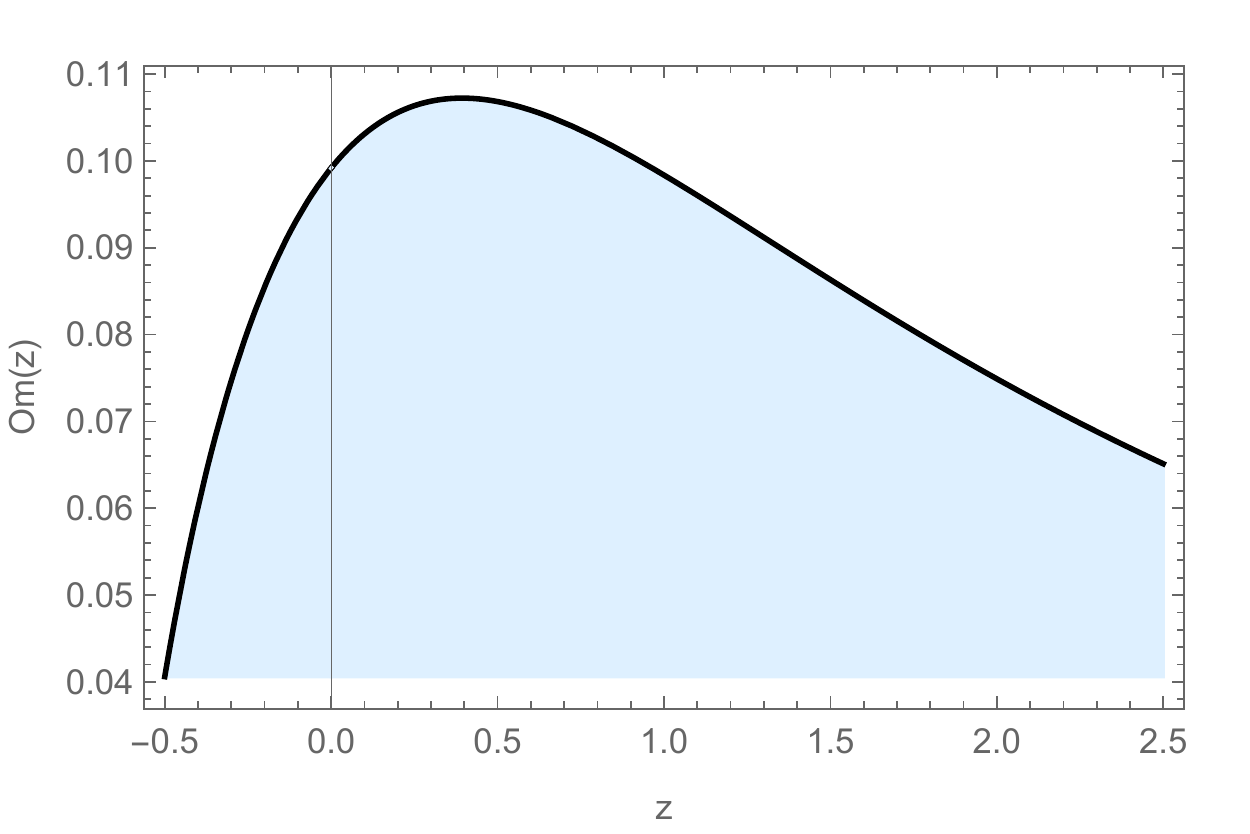}
\endminipage\hfill
\caption{Graphical representation for $Om(z)$ vs. z from Pantheon$^{+}$ data.} \label{FIG.7}
\end{figure}
The slope of $Om(z)$ can distinguish between different dark energy models, even if the value of the matter density is not known. A positive slope of $Om(z)$ for the dark energy model indicates the phantom $(\omega<-1)$ behaviour whereas the negative slope indicates the quintessence $(\omega>-1)$ model. From this diagnostic test we have analysed that the derived model shows the phantom like behaviour at present epoch, similar features we have obtained in the FIG. \ref{FIG.7} shows the phantom like behaviour of the model. 

%%%%%%%%%%%%%%%%%%%%%%%%%%%%%%%%%%
\section{Conclusions} \label{SecVII}
In this paper, we have considered a generalized form of $f(Q,T)$ model as $f(Q,T)=-\lambda_{1} Q^{m}-\lambda_{2} T^2$, where $\lambda_{1}$, $\lambda_{2}$ and $m$ are model parameters. Within the framework of $f(Q,T)$ gravity theory the gravitational field equations were derived by Xu et al. \cite{Xu19}, from the variational principle, and the general relation describing the non conservation of the matter-energy-momentum tensor was identified. We have extended the formalsim of Xu et al. and derived the basic field equations for the generalized form of the functional $f(Q,T)$. The highly non-linear governing equations appears to be exactly solvable only for small values of the model parameters. Eventhough, numerical methods may be adopted to handle highly non-linear equations we prefer to assume small values of the model parameters and solved the evolution equation for the Hubble parameter to get an analytic expression providing a time decreasing Hubble rate. This procedure provides us with simple structure of the equations without destroying the prime Physics involved in it. 

Using  the most recent observational $H(z)$ and Pantheon$^{+}$ data sets, different cosmological parameters and EoS parameter have been constrained through the MCMC method. Using 32 data points and a $\chi^{2}$ minimization strategy, we rebuilt $H(z)$ and distance modulus for observable $H(z)$ values in the redshift range $0.07<z<1.965$. We also looked at the Pantheon$^{+}$ data, which included 1701 data points apparent magnitude measurements. A comparison of the concordance $\Lambda$CDM model with the present one was performed. Since we are interested in a higher power of the non-metricity to describe the observed phenomena, our focus primarily based on the exponent $m$. While the $H(z)$ data set constrained it as $1.17^{+0.25}_{-0.27}$, the Pantheon$^{+}$ data set puts a limit $1.84^{+1}_{-0.93}\pm 0.054$ favouring fractional powers of the non-metricity in between the linear and the quadratic ones. The central values of the EoS parameter as constrained through the adopted methods become $\omega= -1.11$ and $\omega= -1.09$ for Hubble and pantheon$^{+}$ data sets respectively. It is inferred that the proposed model provides a better fit than the $\Lambda$CDM model (See FIG. \ref{FIG.3}). \\

 Now to check the present scenario of the Universe,  behavior of cosmographic parameters (deceleration $q$, jerk $j$ and snap $s$ parameters ) has been analyzed. The cosmological model undergoes a transition from deceleration to acceleration phase at a transition redshift $z_{t}=1.11$. Some recent constraints are compatible with the extracted value of the transition redshift. At the present epoch, the transiting Universe derived in the model has deceleration parameters $q_{0}=-0.79$. We recreated the cosmographic parameters and it should be noted that a divergence from the $\Lambda$CDM model could indicate that the dark energy and dark matter components interact. The $Om(z)$ parameter as a test of our derived models to see if they differ from the concordance $\Lambda$CDM model. The behaviour of the $Om(z)$ parameter in our model indicates a possible phantom field dominated phase at the present epoch. 

%%%%%%%%%%%%%%%%%%%%%%%%%%%%%%
\begin{table}[H]
\caption{$H(z)$ measurements were made using the CC technique, expressed in [$km\, s^{-1} Mpc^{-1}$] units, along with the corresponding errors.} % title of Table
\centering % used for centering table
\begin{tabular}{c c c c c | c c c c c | c c c c c} % centered columns (10 columns)
\hline\hline %inserts double horizontal lines
No. & $z_{i}$ & $H(z)$ & $\sigma_{H(z)}$ & Ref. & No. & $z_{i}$ & $H(z)$ & $\sigma_{H(z)}$ & Ref. &  No. & $z_{i}$ & $H(z)$ & $\sigma_{H(z)}$ & Ref.\\ [0.5ex] % inserts table
%heading
\hline % inserts single horizontal line
1. & 0.070 & 69.00 & 19.6 & \cite{Zhang14} & 12. & 0.400 & 95.00 & 17.00 &  \cite{Simon05} & 23. & 0.875 & 125.00 & 17.00 &  \cite{Moresco12}\\
2. & 0.090 & 69.00 & 12.0 & \cite{Simon05} & 13. & 0.4004 & 77.00 & 10.20 &  \cite{Moresco16} & 24. & 0.880 & 90.00 & 40.00 & \cite{Stern10}\\
3. & 0.120 & 68.60 & 26.2 & \cite{Zhang14} & 14. & 0.425 & 87.10 & 11.20 &  \cite{Moresco16}  & 25. & 0.900 & 117.00 & 23.00 &  \cite{Simon05}\\
4. & 0.170 & 83.00 & 8.00 &  \cite{Simon05}  & 15. & 0.445 & 92.80 & 12.90 &  \cite{Moresco16}  & 26. & 1.037 & 154.0 & 20.00 &  \cite{Moresco12}\\
5. & 0.179 & 75.00 & 4.00 & \cite{Moresco12} & 16. & 0.47 & 89.00 & 49.60 &  \cite{Ratsimbazafy17}  & 27. & 1.300 & 168.0 & 17.00 &  \cite{Simon05} \\
6. & 0.199 & 75.00 & 5.00 &  \cite{Moresco12} & 17. & 0.4783 & 80.90 & 9.00 &  \cite{Moresco16}  & 28. & 1.363 & 160.00 & 33.60 &  \cite{Moresco15} \\
7. & 0.200 & 72.90 & 29.60 &  \cite{Zhang14} & 18. & 0.48 & 97.00 & 62.00 &  \cite{Stern10}  & 29. & 1.430 & 177.0 & 18.00 &  \cite{Simon05} \\
8. & 0.270 & 77.00 & 14.00 &  \cite{Simon05}  & 19. & 0.593 & 104.00 & 13.00 &  \cite{Moresco12} & 30. & 1.530 & 140.0 & 14.00 &  \cite{Simon05}\\
9. & 0.280 & 88.80 & 36.60 & \cite{Zhang14} & 20. & 0.680 & 92.00 & 8.00 &  \cite{Moresco12} & 31. & 1.750 & 202.0 & 40.00 & \cite{Simon05}\\
10. & 0.352 & 83.00 & 14.00 &  \cite{Moresco12}& 21. & 0.750 & 98.80 & 33.60 &  \cite{Borghi22} & 32. & 1.965 & 186.5 & 50.4 & \cite{Simon05} \\
11. & 0.380 & 83.00 & 13.50 &  \cite{Moresco16} & 22. & 0.781 & 105.00 & 12.00 & \cite{Moresco12} & &  &  &  & \\[0.5ex] % [1ex] adds vertical space
\hline %inserts single line
\end{tabular}
\label{TABLE III} % is used to refer this table in the text
\end{table}
%%%%%%%%%%%%%%%%%%%%%%%%%%%%%%
\section*{Acknowledgement} ASA acknowledges the financial support provided by University Grants Commission (UGC) through Senior Research Fellowship (File No. 16-9 (June 2017)/2018 (NET/CSIR)), to carry out the research work. SKT and BM acknowledge the support of IUCAA, Pune (India) through the visiting associateship program. The authors are thankful to the anonymous reviewer for the constructive comments and suggestions for the improvement of the paper.

\end{document}